\documentclass[11pt,a4paper]{article}
\pdfoutput=1

\usepackage{amsfonts}
\usepackage{amssymb}
\usepackage{mathrsfs}
\usepackage{amsmath,amssymb}
\usepackage{subeqn}
\usepackage{bbm}
\usepackage[bookmarks=true]{hyperref}
\usepackage[nosort]{cite}
\usepackage[bulletsep]{collref}

\usepackage{color}

\ifx\hypersetup\sadfkjashdfkxja\else
\hypersetup{pdftitle={A Connection between Twistors and Superstring Sigma Models on Coset Superspaces}}
\hypersetup{pdfauthor={Martin Wolf}}
\hypersetup{pdfsubject={Mathematical Physics}}
\hypersetup{pdfkeywords={Twistors, Integrability, Lax Connection, Superstring Sigma Models}}
\hypersetup{plainpages=false}
\hypersetup{bookmarksnumbered=true}
\hypersetup{pdfstartview=FitH}
\hypersetup{pdfpagemode=None}
\hypersetup{colorlinks=false}
\hypersetup{citebordercolor={.5 1 .5}}
\hypersetup{urlbordercolor={.5 1 1}}
\hypersetup{linkbordercolor={1 .7 .7}}
\fi

\makeatletter\renewcommand{\section}{\@startsection
{section}{1}{\z@}{-2.5ex plus -1ex minus
    -.2ex}{2.3ex plus .2ex}{\centering\large\bf\mathversion{bold}}}

\makeatletter\renewcommand{\subsection}{\@startsection{subsection}{2}{\z@}{-3.25ex
plus -1ex minus
   -.2ex}{1.5ex plus .2ex}{\centering\bf\mathversion{bold}}}

\makeatletter\renewcommand{\subsubsection}{\@startsection{subsubsection}{3}{-2.45ex}{-3.25ex
plus -1ex minus -.2ex}{1.5ex plus .2ex}{\centering\bf\mathversion{bold}}}

\makeatletter\renewcommand{\paragraph}{\@startsection{paragraph}{4}{\z@}%
                                    {0.8ex \@plus1ex \@minus.2ex}%
                                    {-.5em}%
                                    {\normalfont\normalsize\bfseries\mathversion{bold}}}

\renewcommand{\thesection}{\arabic{section}.}

\numberwithin{paragraph}{section}
\renewcommand\theparagraph {\S\thesection\@arabic\c@paragraph.\kern-8pt}

\numberwithin{equation}{section}
\renewcommand{\theequation}{\thesection\arabic{equation}}

\renewcommand*\l@section{\@dottedtocline{1}{0em}{1.5em}}
\renewcommand*\l@subsubsection{\@dottedtocline{4}{3.8em}{3.2em}}

\renewcommand\tableofcontents{%
    \section*{\large\contentsname
        \@mkboth{%
          \MakeUppercase\contentsname}{\MakeUppercase\contentsname}}%
       {\baselineskip=15pt plus 2pt minus 1pt
    \@starttoc{toc}}%
}

\renewenvironment{thebibliography}[1]
     {\section*{\large\centering{\refname}
        \@mkboth{\MakeUppercase\refname}{\MakeUppercase\refname}}%
     \list{\@biblabel{\@arabic\c@enumiv}}%
           {\settowidth\labelwidth{\@biblabel{#1}}%
            \leftmargin\labelwidth
            \advance\leftmargin\labelsep
            \@openbib@code
            \usecounter{enumiv}%
            \let\p@enumiv\@empty
            \renewcommand\theenumiv{\@arabic\c@enumiv}}%
      \sloppy
      \clubpenalty4000
      \@clubpenalty \clubpenalty
      \widowpenalty4000%
      \sfcode`\.\@m
 \catcode`\^^M=10%
}

\DeclareFontFamily{U}{rsf}{}
\DeclareFontShape{U}{rsf}{m}{n}{
  <5> <6> rsfs5 <7> <8> <9> rsfs7 <10-> rsfs10}{}
\DeclareMathAlphabet\Scr{U}{rsf}{m}{n}

\newcommand{\dbar}{\bar\partial}
\newcommand{\dd}{\mathrm{d}}
\newcommand{\di}{\mathrm{i}}

\newcommand{\ewith}{\quad\mbox{with}\quad}
\newcommand{\eand}{\quad\mbox{and}\quad}
\newcommand{\efor}{\quad\mbox{for}\quad}

\newcommand{\IC}{\mathbbm{C}}
\newcommand{\IR}{\mathbbm{R}}
\newcommand{\IZ}{\mathbbm{Z}}
\newcommand{\IP}{\mathbbm{P}}

\newcommand{\cP}{\mathscr{P}}

\newcommand{\CA}{\mathcal{A}}
\newcommand{\CL}{\mathcal{L}}
\newcommand{\CN}{\mathcal{N}}
\newcommand{\CM}{\mathcal{M}}
\newcommand{\CO}{\mathcal{O}}
\newcommand{\CU}{\mathcal{U}}
\newcommand{\CP}{\mathcal{P}}
\newcommand{\CF}{\mathcal{F}}
\newcommand{\CS}{\mathcal{S}}
\newcommand{\CD}{\mathcal{D}}
\newcommand{\CE}{\mathcal{E}}
\newcommand{\CH}{\mathcal{H}}

\newcommand{\sSL}{\mathsf{SL}}
\newcommand{\sGL}{\mathsf{GL}}
\newcommand{\sSO}{\mathsf{SO}}
\newcommand{\sU}{\mathsf{U}}

\newcommand{\sPSU}{\mathsf{PSU}}

\newcommand{\sOSp}{\mathsf{OSp}}

\newcommand{\fg}{\mathfrak{g}}
\newcommand{\fh}{\mathfrak{h}}

\newcommand{\da}{{\dot\alpha}}
\newcommand{\db}{{\dot\beta}}
\newcommand{\dc}{{\dot\gamma}}

\begin{document}

\begin{titlepage}

\setcounter{page}{0}
\renewcommand{\thefootnote}{\fnsymbol{footnote}}

\begin{flushright}
DAMTP 2009--52\\[.5cm]
\end{flushright}

\vspace*{1cm}

\begin{center}

{\LARGE\textbf{\mathversion{bold}A Connection between Twistors and\\ Superstring Sigma Models on Coset Superspaces}\par}

\vspace*{1cm}

{\large
 Martin Wolf\footnote{Also at the Wolfson College,
 Barton Road, Cambridge CB3 9BB, United Kingdom.} \footnote{{\it E-mail address:\/}
\href{mailto:m.wolf@damtp.cam.ac.uk}{\ttfamily m.wolf@damtp.cam.ac.uk}
}}

\vspace*{1cm}

{\it Department of Applied Mathematics and Theoretical Physics\\
University of Cambridge\\
Wilberforce Road, Cambridge CB3 0WA, United Kingdom}

\vspace*{1cm}

{\bf Abstract}
\end{center}

\vspace*{-.3cm}

\begin{quote}

We consider superstring sigma models that are based on coset superspaces $G/H$ in which
$H$ arises as the fixed point set of an order-4 automorphism of $G$. We show
by means of twistor theory that the corresponding first-order system, consisting of the
Maurer--Cartan equations and the equations of motion, arises from a
dimensional reduction of some generalised self-dual Yang--Mills equations in
eight dimensions.
Such a relationship might help shed light on the explicit construction of solutions to the
superstring equations including their hidden symmetry structures and thus on the properties of their
gauge theory duals.

\vfill
\noindent
22th July 2009

\end{quote}

\setcounter{footnote}{0}\renewcommand{\thefootnote}{\arabic{thefootnote}}

\end{titlepage}

\tableofcontents

\bigskip
\bigskip
\hrule
\bigskip
\bigskip

\section{Introduction}

Remarkable advancements in our understanding of maximally $\CN=4$ supersymmetric
Yang--Mills (SYM) theory have been made possible due to its integrability in the
planar limit.
This theory appears to be equivalent to  type IIB superstring theory on
AdS$_5\times S^5$ via the AdS/CFT correspondence and in particular at strong
coupling it is described by classical superstrings. 
 In \cite{Bena:2003wd}, the well-known classical 
integrability of the bosonic AdS$_5\times S^5$ string sigma model was shown 
to extend to its $\kappa$-symmetric Green--Schwarz-type fermionic generalisation
\cite{Metsaev:1998it} (see also \cite{Berkovits:1999zq}).
In this formulation, the superstring sigma model action
is based on the coset superspace $\sPSU(2,2|4)/(\sSO(1,4)\times\sSO(5))$,
where the denominator group arises as fixed point set of an order-4 automorphism of
$\sPSU(2,2|4)$. It is this latter feature that allows for the
construction of conserved non-local charges \`a la L\"uscher
\& Pohlmeyer \cite{Luscher:1977rq} for the superstring on
AdS$_5\times S^5$ \cite{Bena:2003wd} (see also \cite{Vallilo:2003nx,Alday:2003zb,Kazakov:2004qf,Beisert:2005bm,Dorey:2006zj}).\footnote{Aspects 
related to involutivity of the charges were discussed in \cite{Dorey:2006mx}.} 
For a discussion of integrability of the superstring model with the gauges 
fixed and the Virasoro constraints imposed, see \cite{Arutyunov:2004yx,Alday:2005gi} and
e.g.~\cite{Chen:2006gea,Hayashi:2007bq,Grigoriev:2007bu,Mikhailov:2007xr,Jevicki:2007aa,Grigoriev:2008jq,Klose:2008rx,Miramontes:2008wt,Hollowood:2009tw,Giombi:2009ms,Alday:2009ga,Alday:2009yn}.

In this work, we consider superstring sigma models that are based
on coset superspaces $G/H$. Even though our analysis can
be extended to more general cases, 
we always assume that $H$
arises as fixed point set of an order-4 automorphism. 
This then includes the above-mentioned case of type IIB superstrings
on AdS$_5\times S^5$. This also includes type IIA superstring theory on
AdS$_4\times \IC P^3$ in a peculiar partial $\kappa$-symmetry gauge with
$G/H=\sOSp(2,2|6)/(\sSO(1,3)\times \sU(3))$ \cite{Arutyunov:2008if,Stefanski:2008ik} (see also \cite{Gomis:2008jt,Grassi:2009yj}). This particular string theory is the gravitational dual
of the 't Hooft limit of a three-dimensional Chern-Simons matter theory that has recently 
been proposed to be the low-energy description of stacks of
M2-branes on $\IR^8/\IZ_k$ \cite{Aharony:2008ug}. 
Notice that it is a quite generic feature that superstring sigma models based on
coset superspaces of the above type are classically integrable and in fact,
this even extends to models on coset 
(super)spaces with order-$k$ automorphisms \cite{Young:2005jv}.

The full system of the corresponding superstring equations 
consists of i) the Maurer--Cartan equations and ii) the equations of motion that
follow upon varying an associated action functional. This 
set of equations is referred to as the first-order system for the superstring.\footnote{Of course,
these equations should be complemented by the Virasoro constraints.}
We shall discuss the integrability of this system from a different 
point of view:
By using twistor methods, we show that the first-order system of the superstring
arises via a dimensional reduction of some 
generalised self-dual Yang--Mills (SDYM) theory in eight dimensions. The
reason for considering eight dimensions lies in the necessity
of having three `Higgs fields' (as a result of
the $\IZ_4$-grading) after the dimensional reduction. Recall that there are
various generalisations of the four-dimensional SDYM equations to $\IR^d$ with
$d>4$ \cite{Corrigan:1982th,Ward:1983zm,Ivanova:1993ws} and some solutions to these
generalised equations
were, for instance, constructed in \cite{Ivanova:1993ws,Fairlie:1984mp,Fubini:1985jm,Popov:1992cx,Ivanova:1992nj,Loginov:2005ns,Loginov:2008tn}.
See also \cite{Corrigan:1984si} for an extension of the ADHM construction \cite{Atiyah:1978ri}. 
Below, we will 
identify the theory that gives rise to the Lax 
formulation of superstring theory on $G/H$. 
Before discussing the superstring case, however, we shall review the case
of symmetric space sigma models thereby setting up our notation and conventions.

Since the present approach is based on twistor theory, one may naturally hope that it will turn out
useful for the construction of explicit solutions to the superstring equations of motion
by e.g.~using twistor methods like Ward's splitting approach \cite{Ward:1977ta} (see also \cite{Ward:1983zm})
and 
for the study of the hidden symmetry structures. This in turn would
shed light on the properties of (strongly coupled) gauge theory via the holographic correspondence.
We will briefly comment on this at the end of this work. 

\section{Symmetric space coset models and self-dual Yang--Mills theory}

\subsection{Symmetric space coset models}

Let $G$ be a Lie group and $H$ a Lie subgroup of $G$ and consider the coset 
$G/H:=\{gH\,|\,g\in G\}$. We shall assume that $H$ arises as the fixed point set 
of an order-2 automorphism of $G$. This means 
that at the Lie algebra level $\fg:=\operatorname{Lie}(G)$ we have a 
$\IZ_2$-decomposition according to $\fg\cong \fg_{(0)} \oplus \fg_{(2)}$,
where $\fg_{(0)}:=\operatorname{Lie}(H)$ and 
\begin{equation}
 [\fg_{(0)},\fg_{(0)}]\ \subset\ \fg_{(0)}~,~~~
 [\fg_{(0)},\fg_{(2)}]\ \subset\ \fg_{(2)}~\eand
 [\fg_{(2)},\fg_{(2)}]\ \subset\ \fg_{(0)}~.
\end{equation}
If these relations are satisfied, $G/H$ is said to be a symmetric space.
In what follows, we will often denote $\fg_{(0)}$ by $\fh$.

To define the sigma model action, we consider a map $g\,:\,\Sigma\to G$, where $\Sigma$ is a
world-sheet surface with a metric of Lorentzian signature $({+}{-})$, and
introduce the flat current
\begin{equation}\label{eq:flatcurrent}
 j\ :=\ g^{-1} \dd g\ =\ j_{(0)} + j_{(2)}\ =\ A +j_{(2)}~,
 \ewith A\ :=\ j_{(0)} \in\ \fh\eand j_{(2)}\ \in\ \fg_{(2)}~.
\end{equation}
 The  dynamical two-dimensional  fields will take values
in the coset space $G/H$. The action that describes them should
 simultaneously be invariant under the global (left) $G$-transformations of the form
\begin{subequations}
\begin{equation}\label{eq:gtrafo}
 g\ \mapsto\ g_0g\efor g_0\ \in\ G~,
\end{equation}
and the local (right) $H$-transformations of
the form
\begin{equation}\label{eq:htrafo}
 g\ \mapsto\ gh\efor h\ \in\ H~.
\end{equation}
\end{subequations}
By construction, the current $j$ is invariant under \eqref{eq:gtrafo}. 
Under \eqref{eq:htrafo}, the $A$-part of $j$ in \eqref{eq:flatcurrent} 
transforms  as a connection, $A\mapsto h^{-1}Ah+h^{-1}\dd h$, while $j_{(0)}$
transforms covariantly, $j_{(0)}\mapsto h^{-1} j_{(0)} h$.

The sigma model action is then given by
\begin{equation} 
 S\ =\ \tfrac{1}{2}\int_\Sigma\mbox{tr}\,\big[j_{(2)}\wedge{*j}_{(2)}\big]~.
\end{equation}
Here, `$*$' is the Hodge star operator on $\Sigma$ and `tr' the
trace on $\fg$ compatible with the $\IZ_2$-grading. If we set 
\begin{equation}\label{eq:DefofNabla}
 \nabla\alpha\ :=\ \dd\alpha+A\wedge\alpha-(-1)^p\alpha\wedge A
\end{equation}
for a Lie algebra-valued $p$-form $\alpha$ on $\Sigma$, then
the corresponding first-order system may be written as 
\begin{equation} \label{eq:1stOrderSys}
 \dd A + A\wedge A  + j_{(2)}\wedge j_{(2)}\ =\ 0~,~~~\nabla  j_{(2)}\ =\ 0\eand
 \nabla{*j}_{(2)}\ =\ 0~, 
\end{equation}
where the first two equations are the $\fh$ and $\fg_{(2)}$ components of the 
Maurer--Cartan
equation. 
As is well-known, the first-order system is equivalent to the flatness,
\begin{subequations}\label{eq:Flatness}
\begin{equation}\label{eq:flatness}
 \dd J(\zeta)+J(\zeta)\wedge J(\zeta)\ =\ 0~,
\end{equation}
of a Lax connection $J(\zeta)$, with $\zeta$ 
a complex spectral parameter:
\begin{equation}\label{eq:JLambda}
 J(\zeta)\ :=\ A + \tfrac12(\zeta+\zeta^{-1})\, j_{(2)} + \tfrac12(\zeta-\zeta^{-1})\, {*j}_{(2)}~.
\end{equation}
To arrive at \eqref{eq:1stOrderSys} from \eqref{eq:Flatness} we note that on
a world-sheet $\Sigma$ with a Lorentzian signature metric we have $**=1$.
We also have
$\alpha\wedge{*\beta}+{*\alpha}\wedge\beta=0$
for two one-forms $\alpha$ and $\beta$ on $\Sigma$.
Notice that the flatness equation \eqref{eq:flatness} follows as compatibility condition
for an auxiliary linear problem
\begin{equation}\label{eq:ALP}
 \big[\dd+J(\zeta)\big]\psi\ =\ 0~,
\end{equation}
\end{subequations}
where $\psi$ is some $G$-valued function that depends on the spectral parameter $\zeta$.

\subsection{Twistors and self-dual Yang--Mills theory}\label{sec:TSDYM}

Let us now explain how the system \eqref{eq:1stOrderSys}, \eqref {eq:Flatness} in conformal gauge
arises from SDYM 
theory in four dimensions. To this end, we start from the twistor approach.
For text-book treatments of SDYM theory in the context of twistor theory, we refer to
\cite{Ward:1990vs,Mason:1991rf}. 

Consider complexified four-dimensional 
space-time $\CM^4:=\IC^4$. We have the identification $T\CM^4\cong \CS\otimes\tilde\CS$, 
where $\CS$ and $\tilde \CS$ are the two spinor bundles of undotted and dotted spinors
on $\CM^4$,
and so we may consider the projective
co-spin bundle $\CF^5:=\IP(\tilde\CS^*)\cong\IC^4\times\IC P^1$ over $\CM^4$. 
We shall refer to $\CF^5$ as correspondence space. The spaces $\CM^4$ and
$\CF^5$ may be coordinatised by  $x^{\alpha\db}$ and 
$(x^{\alpha\db},\lambda_\da)$, where $\lambda_\da$ are homogeneous
coordinates on $\IC P^1$ and $\alpha,\beta,\ldots=1,2$, $\da,\db,\ldots=\dot1,\dot2$.
On the spinor space $\CS$ (and
similarly on $\tilde \CS$) we have a symplectic form 
$\varepsilon_{\alpha\beta}=\varepsilon_{[\alpha\beta]}$
with $\varepsilon_{\alpha\gamma}\varepsilon^{\gamma\beta}={\delta_\alpha}^\beta$ and
$\varepsilon_{12}=-1$,
which can be used to raise and lower spinor indices.
If we let $\partial_{\alpha\db}:=\partial/\partial x^{\alpha\db}$, then we define
the twistor distribution to be the rank-2 distribution $\CD$ on $\CF^5$ given
by 
\begin{equation}\label{eq:TwistorDistribution}
 \CD\ :=\ \operatorname{span}\big\{V_\alpha:=\lambda^\db\partial_{\alpha\db}\big\}~.
\end{equation}
Since $\CD$ is integrable, it defines a foliation of $\CF^5$, the resulting quotient
will be twistor space, a three-dimensional complex manifold denoted by $\CP^3$. We
have thus established the following double fibration:
\begin{equation}\label{eq:DoubleFibration}
 \begin{picture}(50,40)
  \put(0.0,0.0){\makebox(0,0)[c]{$\CP^3$}}
  \put(64.0,0.0){\makebox(0,0)[c]{$\CM^4$}}
  \put(34.0,33.0){\makebox(0,0)[c]{$\CF^5$}}
  \put(7.0,18.0){\makebox(0,0)[c]{$\pi_1$}}
  \put(55.0,18.0){\makebox(0,0)[c]{$\pi_2$}}
  \put(25.0,25.0){\vector(-1,-1){18}}
  \put(37.0,25.0){\vector(1,-1){18}}
 \end{picture}
\end{equation}
where $\pi_2$ is the trivial projection and $\pi_1\,:\,(x^{\alpha\db},\lambda_\da)\mapsto (z^\alpha=x^{\alpha\db}\lambda_\db,\lambda_\da)$. Hence, $\CP^3\subset\IC P^3$ can be identified
with $\CO(1)\otimes\IC^2\to\IC P^1$, where $\CO(m)$ are the homogeneous polynomials of degree
$m$ on $\IC P^1$. Furthermore, a point $x\in\CM^4$ corresponds to a projective
line $\IC P^1_x\hookrightarrow\CP^3$ in twistor space, while a point $(z,\lambda)\in\CP^3$ 
corresponds to a two-dimensional totally null-plane
in space-time $\CM^4$. Such a plane may be parametrised as
$x^{\alpha\db}=x_0^{\alpha\db}+\mu^\alpha\lambda^\db$,
with $x_0^{\alpha\db}=\operatorname{const.}$ and $\mu^\alpha$ arbitrary.

Consider now a rank-$r$ holomorphic vector bundle $\CE\to\CP^3$ and its pull-back 
$\pi^*_1\CE\to\CF^5$.\footnote{One may impose the additional condition of having
a trivial determinant line bundle $\det\CE$ what would reduce the structure group
$\sGL(r,\IC)$ to $\sSL(r,\IC)$.}
Both the twistor space and the correspondence space can be covered by two coordinate patches
which we denote by $\CU_\pm$ and $\hat\CU_\pm$, respectively. Then the bundles $\CE$ and
$\pi^*_1\CE$ are characterised by transition functions $f_{+-}$ on $\CU_+\cap\CU_-$ and
$\pi^*_1f_{+-}$ on $\hat\CU_+\cap\hat\CU_-$. In what follows, we shall not make
a notational
distinction between $f_{+-}$ and $\pi^*_1 f_{+-}$ and simply write $f_{+-}$ for both bundles.
By definition of a pull back, $f_{+-}$ is constant along $\pi_1\,:\,\CF^5\to\CP^3$ and thus
is annihilated by the vector fields of the twistor distribution \eqref{eq:TwistorDistribution}.
Letting $\dbar_\CP$ and $\dbar_\CF$ be the anti-holomorphic parts of the exterior derivatives
on $\CP^3$ and $\CF^5$, respectively, we have $\pi_1^*\dbar_\CP=\dbar_\CF\circ\pi^*_1$. Hence,
the transition function $f_{+-}$ is also annihilated by $\dbar_\CF$. 

We shall also assume that $\CE$ is topologically trivial and
holomorphically trivial when restricted to any 
$\IC P^1_x\hookrightarrow\CP^3$ for $x\in\CM^4$. These conditions then imply 
the existence of smooth $\sGL(r,\IC)$-valued functions $\psi_\pm$ on $\hat\CU_\pm$ such
that $f_{+-}$ can be decomposed as $f_{+-}=\psi_+^{-1}\psi_-$ with $\dbar_\CF\psi_\pm=0$,
i.e.~the $\psi_\pm$ are holomorphic on $\hat\CU_\pm$. Clearly, this splitting is not
unique, since one can always perform the transformation $\psi_\pm\mapsto g\psi_\pm$, where
$g$ is some globally defined $\sGL(r,\IC)$-valued holomorphic function on $\CF^5$ (hence it is constant
on $\IC P^1$). The choice of $g$ will correspond to a choice of gauge for the
Yang--Mills
gauge potential on space-time. Since $V_\alpha^\pm f_{+-}=0$, 
where $V^\pm_\alpha$ are the restrictions of $V_\alpha$ to the coordinate patches $\hat\CU_\pm$,
we find
\begin{equation}\label{eq:splitting}
 \psi_+ V_\alpha^+\psi_+^{-1}\ =\ \psi_- V_\alpha^+\psi_-^{-1}~
\end{equation}
on $\hat\CU_+\cap\hat\CU_-$.
Explicitly, $V^\pm_\alpha=\lambda^\db_\pm\partial_{\alpha\db}$ with $\lambda^+_\da:=\lambda_\da/\lambda_{\dot 1}
=:(1,\lambda_+)^T$ and $\lambda^-_\da:=\lambda_\da/\lambda_{\dot 2}=:(\lambda_-,1)^T$, where
$(x^{\alpha\db},\lambda_\pm)$ are local coordinates on $\hat\CU_\pm$. Therefore, by an
extension of Liouville's theorem, the expressions \eqref{eq:splitting} can be at most
linear in $\lambda_+$ and thus we may introduce a Lie algebra-valued one-form $\CA$
on $\CF^5$
which has components only along $\CD$,
\begin{equation}
 V_\alpha\lrcorner \CA|_{\hat\CU_\pm}\ :=\ \CA_\alpha^\pm\ =\ \psi_\pm V_\alpha^\pm \psi_\pm^{-1}\ =\ 
 \lambda^\db_\pm\CA_{\alpha\db}~,
\end{equation}
where $\CA_{\alpha\db}$ is $\lambda_\pm$-independent.
This can be re-written as
\begin{equation}
 (V^\pm_\alpha+\CA^\pm_\alpha)\psi_\pm\ =\ \lambda^\db_\pm\nabla_{\alpha\db}\psi_\pm\ =\ 0~,
 \ewith
 \nabla_{\alpha\db}\ :=\ \partial_{\alpha\db}+\CA_{\alpha\db}~.
\end{equation}
The compatibility conditions for this linear system read as
\begin{equation}\label{eq:SDYM}
 [\nabla_{\alpha\db},\nabla_{\gamma\dot\delta}]+[\nabla_{\alpha\dot\delta},\nabla_{\gamma\db}]\ =\ 0~,
\end{equation}
and this is nothing but the SDYM equations, since
\begin{equation}
 [\nabla_{\alpha\db},\nabla_{\gamma\dot\delta}]\ =\ \varepsilon_{\alpha\gamma}f_{\db\dot\delta}
 +\varepsilon_{\db\dot\delta}f_{\alpha\beta}~,
\end{equation}
where $f_{\alpha\beta}$ (respectively, $f_{\da\db}$) represents the self-dual (respectively, anti-self-dual)
part of the field strength. 

In summary, we have described a one-to-one correspondence between equivalence classes of 
holomorphic vector bundles\footnote{Recall that the holomorphic vector bundles $\CE$ and
$\pi_1^*\CE$ are defined
up to the equivalence $f_{+-}\sim h_+^{-1}f_{+-}h_-$, where the $h_\pm$ are holomorphic
$\sGL(r,\IC)$-valued function on $\hat\CU_\pm$ and $V^\pm_\alpha h_\pm=0$. Such
changes do not affect $\CA_{\alpha\dot\beta}$. } over twistor space
that are holomorphically trivial on any projective live $\IC P^1_x\hookrightarrow\CP^3$ and gauge
equivalence classes of solutions
to the SDYM equations on $\CM^4$. This is called the Penrose--Ward transform
\cite{Penrose:1977in,Ward:1977ta}.

Let us now introduce a real structure on $\CP^3$
that yields a split signature real slice in $\CM^4$. 
This can be
done by introducing an anti-holomorphic involution $\tau\,:\,\CP^3\to\CP^3$ 
that is given by
\begin{subequations}\label{eq:RS}
\begin{equation}\label{eq:RSa}
 \tau(z^\alpha,\lambda_\da)
 \ :=\ (\bar z^\beta{C_\beta}^\alpha, {C_{\da}}^\db \bar\lambda_{\db})~,
\end{equation}
where bar denotes complex conjugation and\footnote{Note that these are nothing
but the charge conjugation matrices in split signature \cite{Ketov:1992rh}.}
\begin{equation}\label{eq:RSb}
 ({C_\alpha}^\beta)\ :=\ \begin{pmatrix}
                          0 & 1\\ 1 & 0
                         \end{pmatrix}\eand
 ({C_\da}^\db)\ :=\ \begin{pmatrix}
                          0 & 1\\ 1 & 0
                         \end{pmatrix}.
\end{equation}
\end{subequations}
By virtue of the incidence relation $z^\alpha=x^{\alpha\db}\lambda_\db$, we obtain an
induced involution on $\CM^4$,\footnote{We shall use the same notation $\tau$ for the anti-holomorphic involutions induced on the
different manifolds appearing in \eqref{eq:DoubleFibration}.}
\begin{equation}
 \tau(x^{\alpha\db})\ =\ \bar x^{\gamma\dot\delta} {C_{\gamma}}^\alpha {C_{\dot\delta}}^\db~.
\end{equation}
 The set of fixed points $\tau(x)=x$ is given by $x^{1\dot1}=\bar x^{2\dot2}$
and $x^{1\dot2}=\bar x^{1\dot2}$
and defines a split signature space-time $\CM_\tau^4\cong \IR^{2,2}$:
\begin{equation}
 \dd s^2\ =\ -\tfrac12\varepsilon_{\alpha\beta}\varepsilon_{\dc\dot\delta}\dd x^{\alpha\dc}\dd x^{\beta\dot\delta}
 \ =\ -|\dd x^{1\dot1}|^2+|\dd x^{1\dot2}|^2~.
\end{equation}
We may choose the following parametrisation:
\begin{equation}
 x^{1\dot1}\ =\ \bar x^{2\dot2}\ =:\ -(x^3-\di x^2)\eand
 x^{1\dot2}\ =\ \bar x^{2\dot1}\ =:\ (x^4+\di x^1)~
\end{equation}
leading to
\begin{equation}
 \dd s^2\ =\ (\dd x^1)^2-(\dd x^2)^2-(\dd x^3)^2+(\dd x^4)^2~.
\end{equation}
Notice that the involution $\tau$ can be extended to (holomorphic) functions defined
on the manifolds appearing in the double fibration \eqref{eq:DoubleFibration} and hence to
$\CE$ and $\pi_1^*\CE$ yielding real gauge fields (with values 
in some real form $\fg$ of $\mathfrak{gl}(r,\IC)$).
For a detailed account on the real geometries appearing for
\eqref{eq:RS}, we refer to \cite{Popov:2004rb}.

Using the coordinates $x^\mu=(x^1,\ldots,x^4)$ and $\partial_\mu:=\partial/\partial x^\mu$, the SDYM equations \eqref{eq:SDYM} take the more familiar form
\begin{equation}\label{eq:SDYM2}
 \CF_{12}\ =\ -\CF_{34}~,~~~
 \CF_{13}\ =\ \CF_{24}\eand 
 \CF_{14}\ =\ \CF_{23}~,
\end{equation}
with $\CF_{\mu\nu}:=[\nabla_\mu,\nabla_\nu]$ and $\nabla_\mu:=\partial_\mu+\CA_\mu$,
while the linear system on e.g.~$\hat\CU_+$ is given by
\begin{equation}\label{eq:linsys}
\begin{aligned}
 &\kern4.8cm \CL_1\psi\ =\ 0\ =\ \CL_2\psi~,\\
 &\CL_1\ :=\ \lambda(\nabla_3+\di\nabla_2)+(\nabla_4-\di\nabla_1)\eand 
 \CL_2\ :=\ \lambda(\nabla_4+\di\nabla_1)+(\nabla_3-\di\nabla_2)~,
\end{aligned}
\end{equation}
with $\lambda:=\lambda_+$ and $\psi:=\psi_+$.

To make contact with the discussion of the previous section,
let us perform the linear fractional transformation
\begin{equation}\label{eq:LFT}
 \lambda\ =\ \frac{\zeta-\di}{\zeta+\di}~
\end{equation}
upon which the linear system \eqref{eq:linsys} becomes
\begin{equation}\label{eq:linsys2}
\begin{aligned}
 &\kern4.6cm \hat\CL_1\psi\ =\ 0\ =\ \hat\CL_2\psi~,\\
 &\hat \CL_1\ :=\ \zeta(\nabla_3+\nabla_4)+(\nabla_1+\nabla_2)\eand 
 \hat \CL_2\ :=\ \zeta(\nabla_1-\nabla_2)+(\nabla_3-\nabla_4)~.
\end{aligned}
\end{equation}
Of course, also this linear system leads to \eqref{eq:SDYM2}. 
Assuming that the gauge potential $\CA_\mu$ depends only on $x^1$ and $x^2$
together with $\Phi_{1,2}:=\CA_{3,4}$ and
taking the linear combinations $\frac12[\hat\CL_1\pm\zeta^{-1}\hat\CL_2]$, we find
\begin{subequations} \label{eq:HitchinType}
\begin{equation}
 \begin{aligned}
   {\big[\partial_1+\CA_1+\tfrac12(\zeta+\zeta^{-1})\Phi_1+\tfrac12(\zeta-\zeta^{-1})\Phi_2\big]}\psi\ &=\ 0~,\\
   {\big[\partial_2+\CA_2+\tfrac12(\zeta+\zeta^{-1})\Phi_2+\tfrac12(\zeta-\zeta^{-1})\Phi_1\big]}\psi\ &=\ 0~.
 \end{aligned}
\end{equation}
Thus, we arrive at 
\begin{equation}
 \CF_{12}+[\Phi_1,\Phi_2]\ =\ 0~,~~~
 \nabla_1\Phi_2-\nabla_2\Phi_1\ =\ 0\eand
 \nabla_1\Phi_1-\nabla_2\Phi_2\ =\ 0~.
\end{equation}
\end{subequations}
Let us write $\CA=\CA_1\dd x^1+\CA_2\dd x^2$ and $\Phi=\Phi_1\dd x^1+\Phi_2\dd x^2$.
The system \eqref{eq:HitchinType} 
is almost the component form of \eqref{eq:1stOrderSys}, \eqref {eq:Flatness} when 
written in conformal gauge. In fact, assuming that $\fg$ admits a $\IZ_2$-grading as
discussed above, the sigma model equations on $G/H$ arise as a $\IZ_2$-invariant
subsector of \eqref{eq:HitchinType} determined by this grading. If we let $\Omega\,:\,\fg\to\fg$ be 
the $\IZ_2$-automorphism of $\fg$, we may introduce the projectors $\cP_{(0)}:=\frac12(1+\Omega)$
and $\cP_{(2)}:=\frac12(1-\Omega)$ such that $\fh=\cP_{(0)}(\fg)$ and $\fg_{(2)}=\cP_{(2)}(\fg)$.
The configurations $(\CA,\Phi)$ we are interested in 
are then those with $(\CA,\Phi)=(\Omega(\CA),-\Omega(\Phi))$, i.e.~we may
set
$\CA=:A\in\fh$ and $\Phi=:j_{(2)}\in\fg_{(2)}$. Notice that the system
\eqref{eq:HitchinType} was introduced in \cite{Hitchin:1986vp,Hitchin:1990xx}
(see also \cite{Ward:1985gz,Ward:1987jw}). Notice also that one may
study dimensional reductions of the $\CN=4$ SDYM equations on $\IR^{2,2}$ to two 
dimensions as done in \cite{Popov:2007hb} to end up with sigma models for maps
from certain super Riemann surfaces into $G$ or $G/H$.

In summary, the SDYM equations \eqref{eq:SDYM2} for a gauge potential $\CA_\mu$ 
on four-dimensional flat space with a split signature metric
and with $\CA_{1,2}\in\fh$ and 
$\CA_{3,4}\in\fg_{(2)}$ for $\fg\cong\fh\oplus\fg_{(2)}$
will reduce to the first-order system of a coset model $G/H$ in conformal gauge, provided we assume
that $\CA_\mu$ is independent of $x^3$ and $x^4$. 

\section{Superstring sigma models and generalised self-dual Yang--Mills theory}

\subsection{Superstring sigma models}

Let us now examine superstring models that are based on coset superspaces
$G/H$, where the denominator groups arise as the fixed point sets of order-4
automorphisms of some Lie supergroup $G$. At the
Lie algebra level $\fg:=\operatorname{Lie}(G)$ we have ($m,n=0,\ldots,3$)
\begin{equation}\label{eq:algsplit}
 \fg\ \cong\ \bigoplus_{m=0}^3\fg_{(m)}~,
 \ewith\fg_{(0)}\ :=\ \mbox{Lie}(H)
 \eand [\fg_{(m)},\fg_{(n)}\}\ \subset\ \fg_{(m+n~\mbox{\footnotesize{mod}}~4)}~.
\end{equation}
Here, $\fg_{(0)}$ and $\fg_{(2)}$ are generated by bosonic generators while
$\fg_{(1)}$ and $\fg_{(3)}$ by fermionic ones, respectively and $[\cdot,\cdot\}$
denotes the (graded) commutator on $\fg$. As before, we shall also denote $\fg_{(0)}$
by $\fh$.

To write down the superstring action, we  consider $g:\Sigma\to G$, where
$\Sigma$ is a  world-sheet surface
with a Lorentzian signature metric and introduce the current
\begin{equation}\label{eq:FlatCurrentSST}
 j\ :=\ g^{-1}\dd g\ =\ j_{(0)} + j_{(1)}+j_{(2)}+j_{(3)}~,
 \ewith j_{(m)}\ \in\ \fg_{(m)}
\end{equation}
according to the $\IZ_4$-decomposition of $\fg$. We again set $A:=j_{(0)}$.

The superstring action can be written as a sum of kinetic and Wess-Zumino terms
\cite{Metsaev:1998it,Berkovits:1999zq,Roiban:2000yy},
\begin{equation}\label{eq:action1}
 S\ =\ -\tfrac{T}{2}\int_\Sigma\mbox{str}\,\big[j_{(2)}\wedge{*j_{(2)}}+\kappa j_{(1)}\wedge
 j_{(3)}\big]~,
\end{equation}
where $T=\frac{\sqrt{\lambda}}{2\pi}$ is the string tension
and   `str' denotes the supertrace on
$\fg$ compatible with the $\IZ_4$-grading.
The $\kappa$-symmetry condition requires that $\kappa=\pm 1$; in what follows we shall
assume that $\kappa=1$.\footnote{The opposite 
sign choice is related by a parity transformation on $\Sigma$.} 

By starting from the Maurer--Cartan equation for the current
\eqref{eq:FlatCurrentSST}
\begin{equation}\label{eq:MC}
 \dd j+j\wedge j\ =\ 0
\end{equation}
and splitting $j$ according to the $\IZ_4$-grading of the algebra, we find
\begin{subequations}\label{eq:1stOrderSST}
\begin{equation}\label{eq:MCDecomp}
 \begin{aligned}
  \dd A+A\wedge A +j_{(1)}\wedge j_{(3)}+j_{(2)}\wedge j_{(2)}+j_{(3)}\wedge j_{(1)}\ &=\ 0~,\\
  \nabla j_{(1)}+j_{(2)}\wedge j_{(3)}+j_{(3)}\wedge j_{(2)}\ &=\ 0~,\\
  \nabla j_{(2)}+j_{(1)}\wedge j_{(1)}+j_{(3)}\wedge j_{(3)}\ &=\ 0~,\\
  \nabla j_{(3)}+j_{(1)}\wedge j_{(2)}+j_{(2)}\wedge j_{(1)}\ &=\ 0~,
 \end{aligned}
\end{equation}
and where we used \eqref{eq:DefofNabla}.
The variation of \eqref{eq:action1} over $g$  together with \eqref{eq:MCDecomp} then yields
the following field equations:
\begin{equation}\label{eq:EoMSST}
 \begin{aligned}
  \nabla{*j_{(2)}}+ j_{(3)}\wedge j_{(3)}-j_{(1)}\wedge j_{(1)}\ &=\ 0~,\\
  j_{(2)}\wedge(j_{(1)}+ {*j_{(1)}})+(j_{(1)}+ {*j_{(1)}})\wedge j_{(2)}\ &=\ 0~,\\
  j_{(2)}\wedge(j_{(3)}- {*j_{(3)}})+(j_{(3)}- {*j_{(3)}})\wedge j_{(2)}\ &=\ 0~.
 \end{aligned}
\end{equation}
\end{subequations}
Eqs.~\eqref{eq:1stOrderSST} constitute the full system of the superstring
equations in first-order form, i.e.~the equations for the algebra-valued
one-form $j$. This system is invariant under the bosonic $H$-gauge transformations
and  the fermionic $\kappa$-gauge symmetry.\footnote{It is also invariant under $2d$ reparametrisations.} 

As was shown in \cite{Bena:2003wd} for type IIB superstrings on AdS$_5\times S^5$, the 
$\IZ_4$-grading makes it possible to 
construct one-parameter families of flat currents which in 
turn yield infinitely many non-local conserved charges \`a la L\"uscher
\& Pohlmeyer \cite{Luscher:1977rq}. In fact, this not only true for the
superstring
on AdS$_5\times S^5$ but is a generic feature of models based on cosets with order-4 automorphisms and with
an action of the form \eqref{eq:action1}.\footnote{See \cite{Young:2005jv}
for the extension to $\IZ_k$-graded coset (super)spaces.}
One may verify that the following combination of the components of the current in
\eqref{eq:FlatCurrentSST}
\begin{subequations}\label{eq:LaxSST}
\begin{equation}\label{eq:1PF}
 J(\zeta)\ :=\ A + \zeta^{-1}\,j_{(1)}+\tfrac{1}{2}(\zeta^2+\zeta^{-2})\,j_{(2)}+\zeta\, j_{(3)}
 +\tfrac{1}{2}(\zeta^2-\zeta^{-2})\,{*j_{(2)}}~,
\end{equation}
where $\zeta$ is a complex spectral parameter, satisfies the flatness
condition
\begin{equation} \label{eq:Flat}
 \dd J(\zeta)+J(\zeta)\wedge J(\zeta)\ =\ 0~,
\end{equation}
and vice versa, imposing this flatness condition leads to the  full system  \eqref{eq:1stOrderSST}
of first-order equations for the current $j$.
As before, \eqref{eq:Flat} follows as compatibility condition of an auxiliary linear problem
\begin{equation}\label{eq:ALPSST}
 \big[\dd+J(\zeta)\big]\psi\ =\ 0~,
\end{equation}
\end{subequations}
where $\psi$ is some $G$-valued function that depends on the spectral parameter $\zeta$.

\subsection{Twistors and generalised self-dual Yang--Mills theory}

As we have seen in Sec.~\ref{sec:TSDYM}, symmetric space coset models follow upon dimensionally
reducing the SDYM equations on $\IR^{2,2}$ down to two dimensions. That way
two components of the SDYM field $\CA_\mu$ combine into a Higgs field 
leading to the current $j_{(2)}$. Superstrings
based on coset superspaces as those mentioned above are described by one-parameter
families of flat currents of the form \eqref{eq:1PF}. If we want to understand
the corresponding superstring equations as a dimensional reduction of some self-duality
equations, we in fact need a theory living in eight dimensions since from \eqref{eq:FlatCurrentSST}--\eqref{eq:LaxSST} 
we conclude that we need three Higgs fields that are
represented by $j_{(1)}$, $j_{(2)}$ and $j_{(3)}$. Furthermore, like
for symmetric coset space models, we should consider the self-duality
equations in split signature, i.e.~on  $\IR^{4,4}$. Recall that there are
various generalisations of the SDYM equations to (Euclidean) higher dimensions 
\cite{Corrigan:1982th,Ward:1983zm,Ivanova:1993ws}. We
shall explain which theory leads to the equations
\eqref{eq:1stOrderSST}, \eqref{eq:LaxSST} in two dimensions.

Consider complexified eight-dimensional 
space-time $\CM^8:=\IC^8$. Furthermore, consider two rank-2 complex vector bundles 
$\CS$ and $\tilde \CS$ over $\CM^8$ and make 
the identification $T\CM^8\cong \CS\otimes\odot^3\tilde\CS$, where `$\odot^p$' denotes
the $p$-th symmetric tensor power; see also Eqs.~\eqref{eq:CoordinatesSST}.
This will reduce the rotation group $\sSL(8,\IC)$ to $(\sSL(2,\IC)\times \sSL(2,\IC))/\IZ_2$.\footnote{This
is somewhat in spirit of para-conformal/quaternionic-conformal 
manifolds \cite{Salamon:1982xx,Bailey:1991xx} which are $4k$-dimensional complex
manifolds $\CM$ where one assumes a factorisation of the tangent bundle $T\CM$ 
into one rank-2 complex vector bundle $\CS$ and
one rank-$2k$ complex vector bundle $\CH$, i.e.~$T\CM\cong \CS\otimes \CH$. In this case the rotation
group $\sSL(4k,\IC)$ is reduced to 
$(\sSL(2,\IC)\times \sSL(2k,\IC))/\IZ_2$. In our example, $k=2$ and we assume in addition 
that $\CH$ is given by $\odot^3\tilde\CS$ for some rank-2 complex vector bundle $\tilde\CS$.
Hence, we obtain $(\sSL(2,\IC)\times \sSL(2,\IC))/\IZ_2$.}
As before, we may consider the projectivisation of $\tilde\CS^*$ and introduce
$\CF^9:=\IP(\tilde\CS^*)\cong\IC^8\times\IC P^1$ over $\CM^8$. 
The spaces $\CM^8$ and
$\CF^9$ may be coordinatised by  $x^{\alpha\db_1\db_2\db_3}$ and 
$(x^{\alpha\db_1\db_2\db_3},\lambda_\da)$, where $x^{\alpha\db_1\db_2\db_3}$ is totally symmetric 
in its dotted indices and $\lambda_\da$ are homogeneous
coordinates on $\IC P^1$.
If we introduce $\partial_{\alpha\db\dc\dot\delta}$ with\footnote{To be more concrete, we have 
$\partial_{\alpha\dot1\dot1\dot1}:=\frac{\partial}{\partial x^{\alpha\dot1\dot1\dot1}}$,
$\partial_{\alpha\dot1\dot1\dot2}:=\frac13\frac{\partial}{\partial x^{\alpha\dot1\dot1\dot2}}$,
$\partial_{\alpha\dot1\dot2\dot2}:=\frac13\frac{\partial}{\partial x^{\alpha\dot1\dot2\dot2}}$ and
$\partial_{\alpha\dot2\dot2\dot2}:=\frac{\partial}{\partial x^{\alpha\dot2\dot2\dot2}}$.}
\begin{equation}
\partial_{\alpha\db_1\db_2\db_3}x^{\beta\dc_1\dc_2\dc_3}\ =\ 
{\delta_\alpha}^{\beta}{\delta_{(\db_1}}^{\dc_1}{\delta_{\db_2}}^{\dc_2}{\delta_{\db_3)}}^{\dc_3}~,
\end{equation} 
where parentheses denote normalised symmetrisation,
we then define the twistor distribution to be the rank-2 distribution $\CD$ on $\CF^9$ given
by 
\begin{equation}\label{eq:TwistorDistributionSST}
 \CD\ :=\ \operatorname{span}\big\{V_\alpha:=\lambda^{\db_1}\lambda^{\db_2}\lambda^{\db_3}
  \partial_{\alpha\db_1\db_2\db_3}\big\}~.
\end{equation}
The reason for this choice of the twistor distribution is that we wish to end up we a
Lax {\it pair\/} containing the
eight components of a gauge potential in eight dimensions; for more details see below.

Since $\CD$ is integrable, it defines a foliation of $\CF^9$. The resulting quotient
will be twistor space, a seven-dimensional complex manifold denoted by $\CP^7$,
\begin{equation}\label{eq:DoubleFibrationSST}
 \begin{picture}(50,40)
  \put(0.0,0.0){\makebox(0,0)[c]{$\CP^7$}}
  \put(64.0,0.0){\makebox(0,0)[c]{$\CM^8$}}
  \put(34.0,33.0){\makebox(0,0)[c]{$\CF^9$}}
  \put(7.0,18.0){\makebox(0,0)[c]{$\pi_1$}}
  \put(55.0,18.0){\makebox(0,0)[c]{$\pi_2$}}
  \put(25.0,25.0){\vector(-1,-1){18}}
  \put(37.0,25.0){\vector(1,-1){18}}
 \end{picture}
\end{equation}
where $\pi_2$ is the trivial projection and $\pi_1\,:\,(x^{\alpha\db_1\db_2\db_3},\lambda_\da)\mapsto (z^{\alpha\db_1\db_2}=x^{\alpha\db_1\db_2\db_3}\lambda_{\db_3},\lambda_\da)$. 
Hence, $\CP^7\subset\IC P^7$ is a holomorphic vector bundle over $\IC P^1$ that can best be
understood in terms of its global holomorphic sections $ H^0(\IC P^1,\CP^7)$ 
which are those of $\CO(1)\otimes\IC^6\to\IC P^1$
with the obvious restrictions on the moduli: $H^0(\IC P^1,\CP^7)\subset H^0(\IC P^1,\CO(1)\otimes\IC^6)$ with $z^{\alpha\db_1\db_2}=x^{\alpha\db_1\db_2\db_3}\lambda_{\db_3}$. Notice that
$H^1(\IC P^1,\CP^7)=0$.
Furthermore, a point $x\in\CM^8$ corresponds to a projective
line $\IC P^1_x\hookrightarrow\CP^7$ in twistor space, while a point $(z,\lambda)\in\CP^3$ 
corresponds to a 2-plane inside $\CM^8$ that is parametrised by
$x^{\alpha\db_1\db_2\db_3}=x_0^{\alpha\db_1\db_2\db_3}+
\mu^\alpha\lambda^{\db_1}\lambda^{\db_2}\lambda^{\db_3}$,
with $x_0^{\alpha\db_1\db_2\db_3}=\operatorname{const.}$ and $\mu^\alpha$ arbitrary.

We consider now a rank-$r|s$ holomorphic (super) vector bundle $\CE\to\CP^7$ and 
its pull-back $\pi^*_1\CE\to\CF^9$. Hence, their structure groups are taken to be
$\sGL(r|s,\IC)$.\footnote{We may
additionally assume that the Berezinian line bundle Ber$\,\CE$ is trivial, thus reducing the 
structure group to $\sSL(r|s,\IC)$.}
 Since $\CP^7$ and $\CF^9$ can be covered by two 
patches, $\CU_\pm$ and $\hat\CU_\pm$, these bundles
 are again characterised by transition functions $f_{+-}$. We shall also assume that 
$\CE$ is topologically trivial and holomorphically trivial when restricted to any 
$\IC P^1_x\hookrightarrow\CP^7$ for $x\in\CM^8$. These conditions then again imply 
the existence of smooth $\sGL(r|s,\IC)$-valued functions $\psi_\pm$ on $\hat\CU_\pm$ such
that $f_{+-}$ can be decomposed as 
\begin{equation}\label{eq:SplitF}
 f_{+-}\ =\ \psi_+^{-1}\psi_-~,\ewith\dbar_\CF\psi_\pm\ =\ 0~.
\end{equation}
Since $V_\alpha^\pm f_{+-}=\lambda^{\db_1}_\pm\lambda^{\db_2}_\pm\lambda^{\db_3}_\pm
  \partial_{\alpha\db_1\db_2\db_3}f_{+-}=0$, 
where $V^\pm_\alpha$ are the restrictions of $V_\alpha$ to the coordinate patches $\hat\CU_\pm$,
we find
\begin{equation}\label{eq:splittingSST}
 \psi_+ V_\alpha^+\psi_+^{-1}\ =\ \psi_- V_\alpha^+\psi_-^{-1}~
\end{equation}
on $\hat\CU_+\cap\hat\CU_-$. Thus, we may introduce a Lie algebra-valued one-form $\CA$
on $\CF^9$ which has components only along $\CD$,
\begin{equation}
 V_\alpha\lrcorner \CA|_{\hat\CU_\pm}\ :=\ \CA_\alpha^\pm\ =\ \psi_\pm V_\alpha^\pm \psi_\pm^{-1}\ =\ 
 \lambda^{\db_1}_\pm\lambda^{\db_2}_\pm\lambda^{\db_3}_\pm
  \CA_{\alpha\db_1\db_2\db_3}~,
\end{equation}
where $\CA_{\alpha\db_1\db_2\db_3}$ is $\lambda_\pm$-independent.
This can be re-written as
\begin{subequations}\label{eq:GenSDYMSys}
\begin{equation}\label{eq:LSSST}
 (V^\pm_\alpha+\CA^\pm_\alpha)\psi_\pm\ =\ \lambda^{\db_1}_\pm\lambda^{\db_2}_\pm\lambda^{\db_3}_\pm
\nabla_{\alpha\db_1\db_2\db_3}\psi_\pm\ =\ 0~,
\end{equation}
with $\nabla_{\alpha\db_1\db_2\db_3}:=\partial_{\alpha\db_1\db_2\db_3}+\CA_{\alpha\db_1\db_2\db_3}$.
The compatibility conditions are given by
\begin{equation}\label{eq:GenSDYM}
 \big[\nabla_{\alpha(\db_1\db_2\db_3},\nabla_{\beta\db_4\db_5\db_6)}\big]\ =\ 0~,
\end{equation}
\end{subequations}
i.e.~all dotted indices are symmetrised. Notice that the anti-symmetric tensor product of two vector representations in eight dimensions decomposes under $(\sSL(2,\IC)\times \sSL(2,\IC))/\IZ_2$
as ${\bf 8}\wedge{\bf 8}\cong{\bf 3}\oplus{\bf 15}\oplus{\bf 7}\oplus{\bf 3}$  and the constraints
\eqref{eq:GenSDYM} just imply the vanishing of the ${\bf 7}$-part of the field strength of the gauge potential
$\CA_{\alpha\db_1\db_2\db_3}$. 

In summary, we have a one-to-one correspondence between equivalence classes of holomorphic
vector bundles over $\CP^7$ that are holomorphically trivial along $\IC P^1_x\hookrightarrow\CP^7$
for $x\in\CM^8$ and gauge equivalence classes of solutions to the generalised self-duality equations
\eqref{eq:GenSDYM} on $\CM^8$. The system \eqref{eq:GenSDYMSys} belongs to the
class $B_q$ in Ward's classification scheme \cite{Ward:1983zm}.

We will now explain that \eqref{eq:GenSDYMSys} yields the Lax connection and the first-order system 
for the superstring. To this end,
we introduce a real structure on $\CP^7$
that yields a split signature real slice in $\CM^8$. 
This can be
done in a similar way as \eqref{eq:RS}. In particular,
we consider the involution
\begin{equation}\label{eq:RSSST}
 \tau(z^{\alpha\db_1\db_2},\lambda_\da)
 \ :=\ (\bar z^{\beta\dc_1\dc_2}{C_\beta}^\alpha{C_{\dc_1}}^{\db_1}
  {C_{\dc_2}}^{\db_2}, {C_{\da}}^\db \bar\lambda_{\db})~,
\end{equation}
where the matrices ${C_\alpha}^\beta$ and ${C_\da}^\db$ are the same
as in \eqref{eq:RSb}. We therefore find
\begin{equation}
 \tau(x^{\alpha\db_1\db_2\db_3})\ =\ \bar x^{\beta\dc_1\dc_2\dc_3} {C_{\beta}}^\alpha 
 {C_{\dc_1}}^{\db_1}{C_{\dc_2}}^{\db_2}{C_{\dc_3}}^{\db_3}~.
 \end{equation}
as induced involution on $\CM^8$.
The set of fixed points $\tau(x)=x$ is given by
\begin{equation}
 x^{1\dot1\dot1\dot1}\ =\  \bar x^{2\dot2\dot2\dot2}~,~~~
 x^{1\dot1\dot1\dot2}\ =\  \bar x^{2\dot1\dot2\dot2}~,~~~
 x^{1\dot1\dot2\dot2}\ =\  \bar x^{2\dot1\dot1\dot2}~,~~~
 x^{1\dot2\dot2\dot2}\ =\  \bar x^{2\dot1\dot1\dot1}~
\end{equation}
and defines a split signature space-time $\CM_\tau^8\cong \IR^{4,4}$:
\begin{equation}\label{eq:Metric}
\begin{aligned}
 \dd s^2\ &=\ \tfrac12 \varepsilon_{\alpha\beta}
 \varepsilon_{\db_1\dc_1}\varepsilon_{\db_2\dc_2}\varepsilon_{\db_3\dc_3}
 \dd x^{\alpha\db_1\db_2\db_3} \dd x^{\beta\dc_1\dc_2\dc_3}\\
   &=\ |\dd x^{1\dot1\dot1\dot1}|^2-3|\dd x^{1\dot1\dot1\dot2}|^2+3|\dd x^{1\dot1\dot2\dot2}|^2
       -|\dd x^{1\dot2\dot2\dot2}|^2~.
\end{aligned}
\end{equation}
We then define
\begin{equation}\label{eq:CoordinatesSST}
\begin{aligned}
 x^{1\dot1\dot1\dot1}\ &=\ \bar x^{2\dot2\dot2\dot2}\ =:\ \tfrac18[(x^5-3 x^8)+\di(3x^1-x^3)]~,\\
 x^{1\dot1\dot1\dot2}\ &=\ \bar x^{2\dot1\dot2\dot2}\ =:\ \tfrac18[(x^6+x^7)+\di(x^2+x^4)]~,\\
 x^{1\dot1\dot2\dot2}\ &=\ \bar x^{2\dot1\dot1\dot2}\ =:\ \tfrac18[(x^5+x^8)-\di (x^1+x^3)]~,\\
 x^{1\dot2\dot2\dot2}\ &=\ \bar x^{2\dot1\dot1\dot1}\ =:\ \tfrac18[(x^6-3x^7)-\di (3x^2-x^4)]~
\end{aligned}
\end{equation}
for real $x^\mu$ with $\mu,\nu,\ldots=1,\ldots,8$.
This parametrisation has been chosen with some hindsight and it
will become transparent momentarily. As before, the involution $\tau$ can be
extended to $\CE$ and $\pi^*_1\CE$ to end up with real gauge fields taking values 
in some real form $\fg$ of $\mathfrak{gl}(r|s,\IC)$.

Inverting \eqref{eq:CoordinatesSST}, the linear system \eqref{eq:LSSST} on e.g.~$\hat\CU_+$ is given by 
\begin{equation}\label{eq:LSSST2}
\begin{aligned}
 &\kern5cm \CL_1\psi\ =\ 0\ =\ \CL_2\psi~,\\
 &\CL_1\ :=\ \lambda^3[(\nabla_5-\nabla_8)-\di(\nabla_1-\nabla_3)]
           -\lambda^2[(3\nabla_6+\nabla_7)-\di(\nabla_2+3\nabla_4)]\\
           &\kern2cm+\lambda[(3\nabla_5+\nabla_8)+\di(3\nabla_3+\nabla_1)]
           -[(\nabla_6-\nabla_7)+\di(\nabla_2-\nabla_4)]~,\\
 &\CL_2\ :=\ \lambda^3[(\nabla_6-\nabla_7)-\di(\nabla_2-\nabla_4)]
           -\lambda^2[(3\nabla_5+\nabla_8)-\di(3\nabla_3+\nabla_1)]\\
       &\kern2cm+\lambda[(3\nabla_6+\nabla_7)+\di(\nabla_2+3\nabla_4)]
             -[(\nabla_5-\nabla_8)+\di(\nabla_1-\nabla_3)]~,
\end{aligned}
\end{equation}
with $\lambda:=\lambda_+$, $\psi:=\psi_+$, $\nabla_\mu:=\partial_\mu+\CA_\mu$ and $\partial_\mu:=\partial/\partial x^\mu$. 
It is a straightforward exercise to compute $[\CL_1,\CL_2\}$
to arrive at \eqref{eq:GenSDYM} in the coordinates $x^\mu$. We shall postpone presenting the result
and first perform an additional transformation of the spectral parameter. In light of our
previous discussion, let us again perform the linear fractional transformation \eqref{eq:LFT}.
After some algebraic manipulations, we find that the linear system \eqref{eq:LSSST2} is equivalent
to
\begin{equation}\label{eq:LSSST3}
\begin{aligned}
 &\kern4cm \hat \CL_1\psi\ =\ 0\ =\ \hat \CL_2\psi~,\\
 &\hat \CL_1\ :=\ \zeta^3(\nabla_3+\nabla_4)
           +\zeta^2(\nabla_7+\nabla_8)
           +\zeta(\nabla_1+\nabla_2)+(\nabla_5+\nabla_6)~,\\
 &\hat \CL_2\ :=\ \zeta^3(\nabla_5-\nabla_6)
           -\zeta^2(\nabla_1-\nabla_2)
            -\zeta(\nabla_7-\nabla_8)
             -(\nabla_3-\nabla_4)~.
\end{aligned}
\end{equation}
Of course, both systems \eqref{eq:LSSST2} and \eqref{eq:LSSST3} lead to the same
compatibility conditions, though \eqref{eq:LSSST3} looks much simpler and 
eventually leads us directly to the Lax pair for the superstring. This was 
the reason for the choice \eqref{eq:CoordinatesSST}. 

The compatibility equations are then given by (see also \eqref{eq:GenSDYM})
\begin{equation}\label{eq:GenSDYMCom}
\begin{aligned}
 \CF_{12}+\CF_{34}+\CF_{78}-\CF_{56}\ &=\ 0~,\\
 \CF_{13}-\CF_{24}+\CF_{67}-\CF_{58}\ &=\ 0~,\\
 \CF_{14}-\CF_{23}-\CF_{57}+\CF_{68}\ &=\ 0~,\\
 \CF_{15}+\CF_{18}-\CF_{26}-\CF_{27}+\CF_{38}-\CF_{47}\ &=\ 0~,\\
 \CF_{16}-\CF_{17}-\CF_{25}+\CF_{28}+\CF_{37}-\CF_{48}\ &=\ 0~,\\
 \CF_{35}-\CF_{46}\ &=\ 0~,\\
 \CF_{36}-\CF_{45}\ &=\ 0~,
\end{aligned}
\end{equation}
where $\CF_{\mu\nu}:=[\nabla_\mu,\nabla_\nu\}$.

To make contact with our discussion about superstring sigma models, let us
us assume that $\CA_\mu$ depends only on $x^1$ and $x^2$ and introduce
$\Phi_{1,2}:=\CA_{3,4}$ and
\begin{equation}
 \begin{aligned}
  \Psi_1\ :=\ \tfrac12 (\CA_5+\CA_6+\CA_7-\CA_8)~,~~~
  \Psi_2\ :=\ \tfrac12 (\CA_5+\CA_6-\CA_7+\CA_8)~,\\
  \Sigma_1\ :=\ \tfrac12 (-\CA_5+\CA_6+\CA_7+\CA_8)~,~~~
  \Sigma_2\ :=\ \tfrac12 (\CA_5-\CA_6+\CA_7+\CA_8)~.\kern-.1cm
 \end{aligned}
\end{equation}
Taking the linear combinations $\frac12[\zeta^{-1}\hat \CL_1\pm \zeta^{-2}\hat \CL_2]$,
we find from \eqref{eq:LSSST3}
\begin{subequations}\label{eq:HitchinTypeGeneral}
\begin{equation}\label{eq:HitchinTypeGeneralA}
\begin{aligned}
 \big[\nabla_1+\zeta^{-1}\Psi_1+
         \tfrac12(\zeta^2+\zeta^{-2})\Phi_1+\zeta\Sigma_1+
          \tfrac12(\zeta^2-\zeta^{-2})\Phi_2\big]\psi\ &=\ 0~,\\
 \big[\nabla_2+\zeta^{-1}\Psi_2+
         \tfrac12(\zeta^2+\zeta^{-2})\Phi_2+\zeta\Sigma_2+
          \tfrac12(\zeta^2-\zeta^{-2})\Phi_1\big]\psi\ &=\ 0~.\\
\end{aligned}
\end{equation}
\pagebreak[4]
The compatibility equations of this system are of Hitchin-type
\begin{equation}\label{eq:GH1}
 \begin{aligned}
  \CF_{12}+[\Phi_1,\Phi_2\}+[\Psi_1,\Sigma_2\}+[\Sigma_1,\Psi_2\}\ &=\ 0~,\\
  \nabla_1\Psi_2-\nabla_2\Psi_1+[\Phi_1,\Sigma_2\}+[\Sigma_1,\Phi_2\}\ &=\ 0~,\\
  \nabla_1\Phi_2-\nabla_2\Phi_1+[\Psi_1,\Psi_2\}+[\Sigma_1,\Sigma_2\}\ &=\ 0~,\\
  \nabla_1\Sigma_2-\nabla_2\Sigma_1+[\Phi_1,\Psi_2\}+[\Psi_1,\Phi_2\}\ &=\ 0~,\\
  \nabla_1\Phi_1-\nabla_2\Phi_2-[\Psi_1,\Psi_2\}+[\Sigma_1,\Sigma_2\}\ &=\ 0~,\\
  [\Phi_1,\Psi_1+\Psi_2\}+[\Psi_1+\Psi_2,\Phi_2\}\ &=\ 0~,\\
  [\Phi_1,\Sigma_1-\Sigma_2\}-[\Sigma_1-\Sigma_2,\Phi_2\}\ &=\ 0~,\\
 \end{aligned}
\end{equation}
\end{subequations}
which,
of course, just follow from \eqref{eq:GenSDYMCom} upon assuming that $\CA_\mu$ depends
only on $x^1$ and $x^2$ and using the above definitions of $\Phi$, $\Psi$ and $\Sigma$. As
before, $\Phi=\Phi_1\dd x^1+\Phi_2\dd x^2$ and similarly for the others.
Eqs.~\eqref{eq:HitchinTypeGeneral} represent almost \eqref{eq:1stOrderSST} and \eqref{eq:LaxSST} in conformal gauge.
If we assume that $\fg$ admits a $\IZ_4$-grading, then
the superstring equations arise as a $\IZ_4$-invariant subsector of  \eqref{eq:HitchinTypeGeneral}.
This is analogous to what happened in the symmetric space case.
Let $\Omega\,:\,\fg\to\fg$ be the $\IZ_4$-automorphism of $\fg$. We then may introduce projectors\footnote{For
details on the grading in the case of superstrings on AdS$_5\times S^5$, see e.g.~\cite{Arutyunov:2009ga}.}
(see also Eqs.~\eqref{eq:algsplit}; $\di:=\sqrt{-1}$)
\begin{equation}
 \cP_{(m)}\ :=\ \tfrac14(1+\di^{3m}\Omega+\di^{2m}\Omega^2+\di^m\Omega^3)~,\ewith
 \fg_{(m)}\ =\ \cP_{(m)}(\fg)~,
\end{equation}
projecting onto the $\fg_{(m)}$-components of $\fg$.
The configurations $(\CA,\Psi,\Phi,\Sigma)$ corresponding to the superstring are those which satisfy 
\begin{equation}
(\CA,\Psi,\Phi,\Sigma)\ =\ (\Omega(\CA),\di\Omega(\Psi),-\Omega(\Phi),-\di\Omega(\Sigma))~. 
\end{equation}
Therefore, for such $(\CA,\Psi,\Phi,\Sigma)$ we may relable
$\CA=:A\in\fh$, $\Psi=:j_{(1)}\in\fg_{(1)}$, $\Phi=:j_{(2)}\in\fg_{(2)}$ and
$\Sigma=:j_{(3)}\in\fg_{(3)}$ eventually arriving at \eqref{eq:1stOrderSST} and \eqref{eq:LaxSST}.

In summary, the first-order system \eqref{eq:1stOrderSST} 
 of the superstring based on a coset superspace $G/H$ with the above properties
can be obtained as a dimensional reduction of the generalised self-duality type equations
\eqref{eq:GenSDYM}, \eqref{eq:GenSDYMCom} for a gauge potential $\CA_\mu$ with the assumptions
 $\CA_{1,2}\in\fh$, $\CA_{3,4}\in\fg_{(2)}$,
$\pm \CA_5+\CA_6+\CA_7\mp \CA_8\in\fg_{(1)}$ and $\CA_5\pm \CA_6\mp \CA_7+\CA_8\in\fg_{(3)}$, where
$\fg\cong\fh\oplus\fg_{(1)}\oplus\fg_{(2)}\oplus\fg_{(3)}$.
As we have dicussed above, gauge equivalence classes of solutions to the 
self-duality type equations \eqref{eq:GenSDYM} are in one-to-one
correspondence with equivalence classes of holomorphic vector bundles over the
twistor space $\CP^7$ and its correspondence space $\CF^9$ which are subject to certain
algebraic constraints.
 Therefore, all solutions to the superstring equations \eqref{eq:1stOrderSST} 
are encoded in these 
holomorphic vector bundles. Of course, physical solutions to \eqref{eq:1stOrderSST}
should additionally obey the Virasoro constraints putting therefore further assumptions
on the admissible vector bundles.

\subsection{Remarks}

\noindent
{\bf Remark 1.}
For the sake of concreteness, we only considered a dimensional reduction leading to a Lorentzian
world-sheet. Of course, one could perform the reduction differently to end up
with a 
Euclidean world-sheet.
Furthermore, as we wanted to re-produce the superstring equations, we considered an anti-holomorphic 
involution \eqref{eq:RSSST} on $\CP^7$ corresponding to a split signature space-time $\IR^{4,4}$. 
One may instead consider an involution on $\CP^7$ leading to a Euclidean
signature real slice in $\CM^8$ and then try to repeat the above procedure to arrive 
at a direct generalisation of the Hitchin equations given in \cite{Hitchin:1986vp}. 
In view of that notice that the Hitchin equations are a key
ingredient in recent constructions of strong coupling gluon scattering amplitudes
in planar $\CN=4$ SYM theory \cite{Alday:2009ga,Alday:2009yn} via the AdS/CFT 
correspondence (see also \cite{Alday:2007hr}).
It would be interesting to see whether these generalised equations would play a role when
extending the results of \cite{Alday:2009ga,Alday:2009yn} to the full background
geometry.

\bigskip
\noindent
{\bf Remark 2.}
Finally, let us make a few comments on non-local charges and hidden symmetry structures.
The above twistor description allows for a geometric re-interpretation of the conserved
non-local charges for the superstring: In fact, these charges follow from
the function $\psi$ appearing in \eqref{eq:ALPSST} for
some particular choice of contour in 
the world-sheet surface upon expanding it in powers of the spectral parameter
\cite{Bena:2003wd}. 
We have just seen how this function is related  
to the transition functions of the holomorphic vector bundles $\CE$ and $\pi^*_1\CE$ 
over the twistor space $\CP^7$ and its correspondence space $\CF^9$. In this respect,
notice that upon reducing \eqref{eq:LSSST2} to two dimensions and making use of 
the definitions of $\Psi$, $\Phi$ and $\Sigma$, one obtains a linear system
equivalent to \eqref{eq:HitchinTypeGeneralA}.
That way one may then study infinitesimal deformations of these vector bundles to
describe the hidden symmetry algebras for the superstring and furthermore, to give an interpretation
of the symmetries in terms of sheaf cohomology along the lines presented in the works \cite{Mason:1991rf,Popov:1998pc,Wolf:2004hp,Popov:2005uv,Wolf:2005sd,Wolf:2006me,Popov:2006qu}.
For example, any deformation algebra of the transition functions of Lie algebra-type
can be mapped into a symmetry algebra for the gauge potential (modulo gauge equivalence); see \cite{Wolf:2005sd,Wolf:2006me} for more details.
This twistor re-interpretation may in turn help shed light on the hidden symmetry structures
appearing in the gauge theory duals via the holographic correspondence.
In view of this it would, for example, be interesting to study the recently uncovered dual
(super)conformal symmetry \cite{Drummond:2006rz,Drummond:2007cf,Drummond:2007au,Drummond:2008vq,Brandhuber:2008pf} (see also \cite{Drummond:2009fd}) in
$\CN=4$ SYM theory and superstring theory on AdS$_5\times S^5$
\cite{Alday:2007hr,Ricci:2007eq,Berkovits:2008ic,Beisert:2008iq,Beisert:2009cs}
in terms of the twistor approach presented here.\footnote{For a recent review
on the Wilson loop/gauge theory scattering amplitude correspondence, see \cite{Alday:2008yw}.} 
Recall that on the string side,
the appearance of the dual superconformal symmetry is due to the `T-self-duality'  
of the superstring sigma model under a certain combination of bosonic and fermionic T-dualities
in Poincar\'e parametrisation \cite{Berkovits:2008ic,Beisert:2008iq}.
A way of interpreting this T-duality is then as a dressing transformation on the space of solutions,
like a B\"acklund transformation (see, e.g.~\cite{Arutyunov:2005nk}).

\bigskip
\bigskip
\noindent
{\bf Acknowledgements.}
I am very grateful to J.~Bedford, N.~Bouatta, N.~Dorey, M.~Dunajski, R.~Ricci, C.~S{\"a}mann and A.~Tseytlin
for important discussions, questions and suggestions.
This work was supported by an STFC Postdoctoral Fellowship and by a Senior Research
Fellowship at the Wolfson College, Cambridge, U.K.

\end{document}